\documentclass[reprint,superscriptaddress,amsmath,amssymb,aps,prl,longbibliography
]{revtex4-1}

\usepackage[T1]{fontenc}
\usepackage[utf8]{inputenc}
\usepackage[unicode=true,breaklinks=true,colorlinks=true]{hyperref}
\hypersetup{citecolor=blue,urlcolor=blue}
\usepackage{graphicx}
\usepackage{bm}
\usepackage{bbold}

\usepackage{mathtools}
\usepackage{tensor}

\begin{document}
\title{Quantum sensing with a single-qubit pseudo-Hermitian system}
\author{Yaoming Chu}
\author{Yu Liu}
\author{Haibin Liu}
\affiliation{School of Physics, International Joint Laboratory on Quantum Sensing and Quantum Metrology, Huazhong University of Science and Technology, Wuhan 430074, China}
\author{Jianming Cai}
\email{jianmingcai@hust.edu.cn}
\affiliation{School of Physics, International Joint Laboratory on Quantum Sensing and Quantum Metrology, Huazhong University of Science and Technology, Wuhan 430074, China}
\date{\today}
	
\begin{abstract}
Quantum sensing exploits fundamental features of quantum system to achieve highly efficient measurement of physical quantities. Here, we propose a strategy to realize a single-qubit pseudo-Hermitian sensor from a dilated two-qubit Hermitian system. The pseudo-Hermitian sensor exhibits divergent susceptibility in dynamical evolution that does not necessarily involve exceptional point. We demonstrate its potential advantages to overcome noises that cannot be averaged out by repetitive measurements. The proposal is feasible with the state-of-art experimental capability in a variety of qubit systems, and represents a step towards the application of non-Hermitian physics in quantum sensing.
\end{abstract}
	%\pacs{Valid PACS appear here}
	%\keywords{Suggested keywords}
	
    \maketitle
	\textit{Introduction.---} Non-Hermitian Hamiltonians usually have complex energy eigenvalues and do not conserve probabilities, thus presumably only serve as phenomenological descriptions of open quantum system \cite{Feshbach1954,Barton1963}. Remarkably, a non-Hermitian Hamiltonian $H$ with an exact $\mathcal{PT}$-symmetry \cite{Bender1998,Dorey2001ii,Bender2002} and more general pseudo-Hermiticity (i.e. $\eta H =H^{\dagger} \eta$ with a Hermitian invertible linear operator $\eta$) \cite{Pauli1943,Lee1969,Mostafazadeh2002ii,Bender2007} can have real spectrum. The discovery has opened a new avenue to intriguing non-Hermitian physics in both classical and quantum systems \cite{Ruter2010,Peng2014,ZJing2018,Hang2013,Zhang2016,Bender2013,Bitter2012,Gao2015,Fleury2015,Choi2018,Tang2016,Xiao2017,Xiao2018,Ganainy2018,Miri2019}. In particular, the concepts of exceptional point \cite{Kato1966,Berry2004,Heiss2012} and $\mathcal{PT}$-phase transition \cite{Guo2009,Li2019} lead to important experimental observations such as single-mode lasers \cite{Feng2014,Hodaei2014}, non-reciprocal light transport \cite{Feng2011,Yin2013}, non-Hermitian topological light steering \cite{Zhao2019},  asymmetric mode switching \cite{Doppler2016} and topological energy transfer \cite{Xu2016} when encircling an exceptional point.
Among a number of important applications, enhancing the sensitivity of energy splitting detection by using exceptional points attracts increasingly intensive theoretical interest \cite{Wiersig2014,Wiersig2016,Liu2016,Ren2017,Sunada2017,Zhang2018}. The unique feature of exceptional point based sensing is the divergent eigenvalue susceptibility arising from the square-root frequency topology around exceptional points \cite{Heiss2012}. Exceptional point based sensing has been demonstrated in $\mathcal{PT}$-symmetric linear coupled-mode optic systems, with potential applications for single-particle detection \cite{Chen2017,Hodaei2017} and optical gyroscope \cite{Lai2019}. Nevertheless, the singular behavior of the energy splitting is counteracted by the eigenstate coalescence \cite{Langbein2018,Lau2018,Chen2019}. Quantum sensing based on non-Hermitian qubit system without involving exceptional points and its potential advantages remain largely unexplored.  
In this work, we propose a novel strategy to harness a qubit probe with pseudo-Hermiticity as a resource for quantum sensing by introducing an additional qubit ancilla \cite{Gunther2008,Kawabata2017,Huang2019,Wu2019}.  The dilated two-qubit Hermitian system, when exposed to a parameter-dependent weak field, reproduces an effective pseudo-Hermitian qubit sensor by postselection, inheriting the influence of the parameter. The eigenstate coalescence without involving exceptional point induces divergent susceptibility in the normalized population of the pseudo-Hermitian qubit sensor, which offers advantages for quantum sensing under realistic noises that cannot be averaged out by repetitive measurements  \cite{Feizpour2011,Jordan2014,Hosten2008,Dixon2009,Starling2009}. The experiment demonstration is feasible in trapped-ion systems and other physical systems.
\begin{figure}[h]
\centering
\includegraphics[width=90mm]{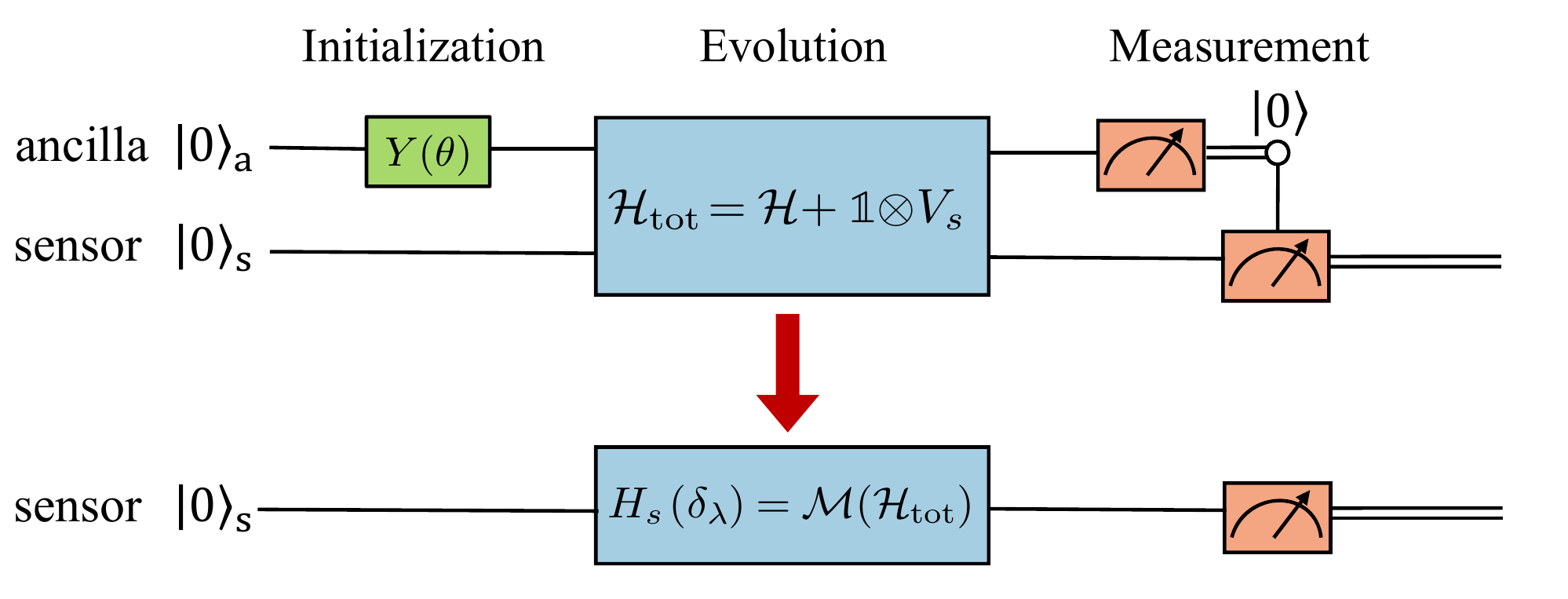}
\caption{The quantum circuit representation of our pseudo-Hermitian quantum sensing protocol. $Y(\theta)$ denotes a rotation of the ancilla qubit around $\hat{y}$ axis by an angle $\theta$ to prepare the initial state of the two-qubit dilated system. The two-qubit dilation system evolves under the Hamiltonian $\mathcal{H}_{\text{tot}}$. The  effective dynamics of the sensor system conditioned on the postselection measurement of the ancilla qubit is described by an effective pseudo-Hermitian Hamiltonian $H_s(\delta_\lambda)$. The outcome of the measurement on the sensor system reveals the information about the parameter $\lambda$. The coalescence of the pseudo-Hermitian $H_s(\delta_\lambda)$ provides an advantageous resource for quantum sensing.
}\label{fig:model}
\end{figure} 

\textit{Basic principle.---} While our proposal is completely general for a generic pseudo-Hermitian Hamiltonian with eigenstate coalescence \cite{supplement}, we start from a simple two-level system with a pseudo-Hermitian Hamiltonian that is described as
\begin{equation}
\label{eq:HP}
 H_{s}(\mathcal{E},\delta)=\mathcal{E}\left[\begin{array}{cc}{0}&{\delta^{-1}}\\{\delta} & {0}\end{array}\right],
\end{equation}
with real eigenvalues $\pm \mathcal{E}$. The associated eigenstates $|\phi_\pm\rangle\propto |0\rangle\pm\delta|1\rangle$ symmetrically coalesce to $|0\rangle$ as $\vert\delta\vert \rightarrow 0$.  The above non-Hermitian Hamiltonian cannot be directly realized in a closed quantum system \cite{Lee2014,Wu2019}. In order to achieve an effective pseudo-Hermitian quantum system satisfying the following Schr\"odinger equation
    \begin{equation}
          i\partial_t\left|\psi(t)\right\rangle_s=H_s\left|\psi(t)\right\rangle_s, 
    \end{equation}
we adopt a Naimark dilation approach \cite{Tang2016,Xiao2018,Wu2019} using an extended Hermitian system. The evolution of the total system (including the system $s$ and the ancilla qubit $a$) $\left|\Psi(t)\right\rangle=\left|0\right\rangle_a\otimes\left|\psi(t)\right\rangle_s+\left|1\right\rangle_a\otimes\left|\chi(t)\right\rangle_s$ follows  a dilated Hamiltonian $\mathcal{H}$ \cite{supplement}. The Naimark dilation approach implements a map $\mathcal{M}:\mathcal{H}\mapsto H_s$, which realizes an effective pseudo-Hermitian dynamics $\left|\psi(t)\right\rangle_s$ of the system $s$ conditioned on the ancilla qubit state $|0\rangle_a$ from a two-qubit Hermitian Hamiltonian \cite{Kawabata2017}. We note that $|\chi(t)\rangle_s=\eta|\psi(t)\rangle_s$ with $\eta$ a positive Hermitian operator fulfilling the condition $(\eta^2+\mathbb{1}) H_s=H_s^{\dagger}(\eta^2+\mathbb{1})$ to guarantee the Hermiticity of the dilated Hamiltonian $\mathcal{H}$. 
        
Our protocol uses a two-qubit dilated Hermitian system to construct a pseudo-Hermitian quantum probe. The main idea is represented by a quantum circuit in Fig.\ref{fig:model}. Without loss of generality, we consider the detection of a weak field acting on the sensor system that is described in the form of $V_s= \lambda\sigma_{x}^{(s)}$, where $\lambda$ is the parameter to be estimated. The full two-qubit dilated system is  described by the following total Hamiltonian as
    \begin{equation}
        \label{eq:dilated_H}
        \mathcal{H}_{\text{tot}}=\mathcal{H}+\mathbb{1}\otimes V_s,
    \end{equation}
with $\mathcal{H}=b \mathbb{1}^{(a)} \otimes \sigma_{x}^{(s)}-c\sigma_{y}^{(a)} \otimes \sigma_{y}^{(s)}$ satisfying $\mathcal{M}(\mathcal{H})=H_s(\delta)$, which is the effective Hamiltonian of the pseudo-Hermitian sensor in Eq.\eqref{eq:HP}. 
As an example to demonstrate the essential idea of our proposal, we choose $\delta=\sqrt{\varepsilon/(1+\varepsilon)}$ and $\mathcal{E}=2\omega\sqrt{\varepsilon(1+\varepsilon)}$, and write the Hamiltonian Eq.(\ref{eq:HP}) in the following familiar form
\begin{equation}
H_s(\omega, \varepsilon)=2\omega\left(\begin{array}{cc}
0    & 1+\varepsilon \\
\varepsilon   & 0
\end{array}\right).
\end{equation}
In the presence of an external weak field $\lambda$, we follow the Naimark dilation approach to derive the effective Hamiltonian for the system. By defining $b_\lambda=4 \omega\varepsilon(1+\varepsilon)/(1+2 \varepsilon)+\lambda$ and $c=2\omega \sqrt{\varepsilon(1+\varepsilon)}/(1+2 \varepsilon)$, we find that $|\chi\rangle_s=\hat{\mathcal{\zeta}}_s|\psi\rangle_s$ with $\hat{\zeta}_s=\text{diag}\left[\kappa, (c+\kappa b_\lambda)/(b_\lambda-\kappa c)\right]$, where $\kappa$ is a parameter depending on the initial condition \cite{supplement}. For simplicity, we choose the parameter $\kappa=\delta$, and the dilated two-qubit system is initialized in the state $|\Psi(0)\rangle\propto \left(\vert0\rangle_a+\delta\vert 1\rangle_a\right)\otimes\vert 0\rangle_s$ by a rotation of the ancilla qubit around $\hat{y}$ axis by an angle $\theta=2\arcsin (\delta / \sqrt{1+\delta^2})$ which is denoted as $Y(\theta)$, see Fig.\ref{fig:model}. In this case, the effective Hamiltonian governing the dynamics of the sensor system's state $|\psi(t)\rangle_s$ is found to be $H_s\left(\delta_\lambda\right)=\mathcal{M}\left(\mathcal{H}_{\text{tot}}\right)$ \cite{supplement} with the eigenvalues $\pm \mathcal{E}_{\lambda}$, wherein
    \begin{equation}
        \begin{aligned}
            & \mathcal{E}_\lambda=\sqrt{b_\lambda^2+c^2},\\
            & \delta_\lambda =(\lambda+2\varepsilon\omega)/\mathcal{E}_\lambda.            
        \end{aligned}
        \label{eq:parameter}
    \end{equation}
    \begin{figure}[t]
    \centering
    \includegraphics[width=86mm]{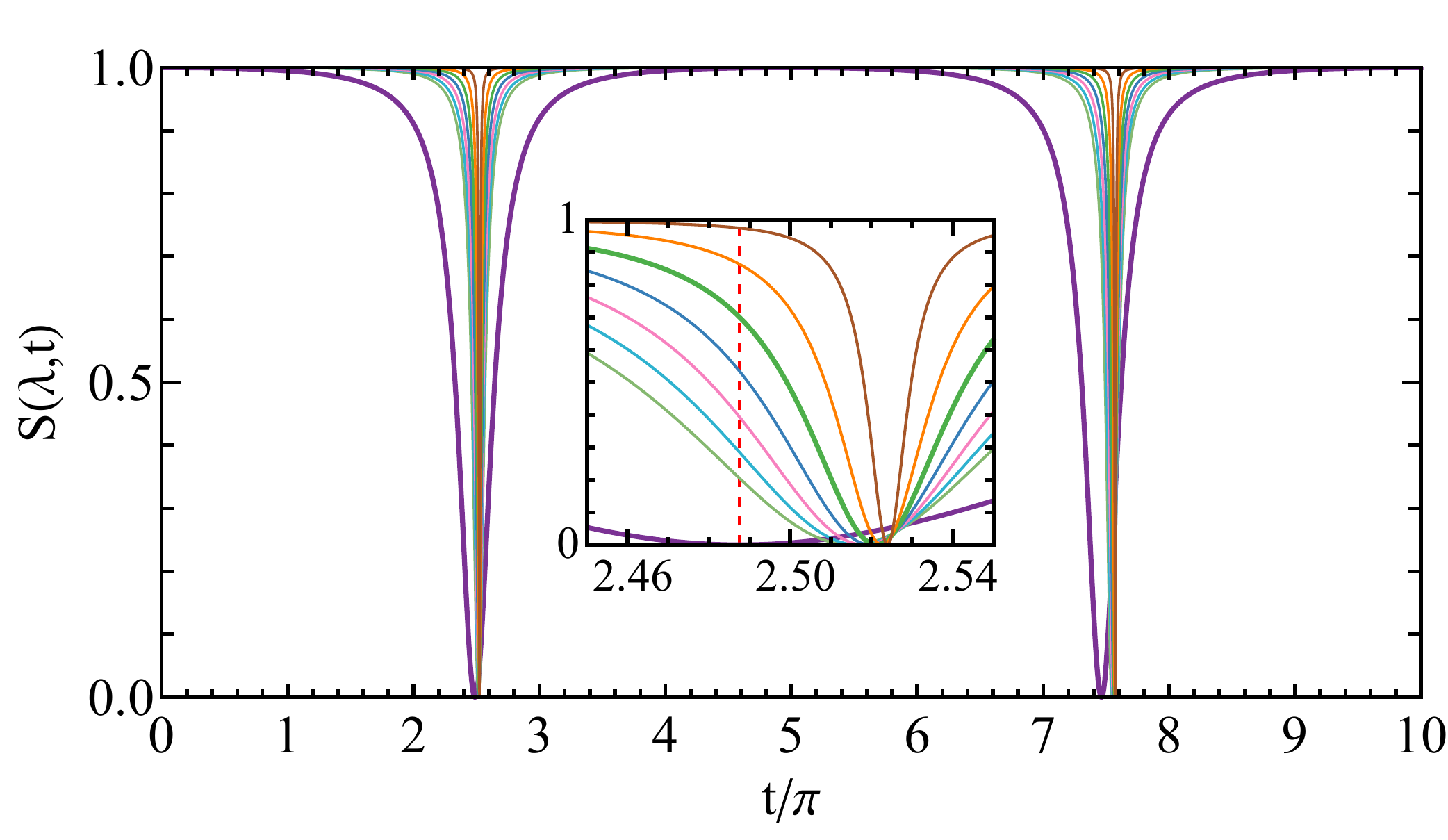}
    \caption{The evolution of the normalized state population of the pseudo-Hermitian sensor system $S(\lambda,t)$ (see Eq.\eqref{eq:parameter} and Eq.\eqref{eq:SB}) as a function of time.
The population exhibits sharp dips at time $t_j=(j-1/2)\pi/\mathcal{E}_\lambda$ ($j\in\mathbb{Z}^{+}$) with a narrow width $\Delta t=2\sqrt{\mathcal{D}_\lambda} /\mathcal{E}_\lambda$. The comparison of the results from different values of $\lambda$, including $\lambda=0$ (purple) and $(\lambda/\varepsilon\omega)+2=(0.072,0.172,0.272,0.372,0.472,0.572,0.672)$ 
(from top to bottom), demonstrate that both the location and the width of the dips are dependent on the parameter $\lambda$. The sharp dip in combination with the $\lambda$-dependent dip location and dip width lead to the feature that  $S(\lambda,t)$ becomes extremely sensitive to the parameter $\lambda$ when the evolution time is set close to $t_j\vert_{\lambda=0}$. The inset shows a zoom in view around $t_1\vert_{\lambda=0}$ (red dashed line). The other parameters are $\varepsilon=0.01$, $\omega=1$.}\label{fig:signal}
    \end{figure}

It can be seen that the effective pseudo-Hermitian Hamiltonian incorporates the influence of the parameter $\lambda$ that is to be estimated. After an evolution time $t$, the state of the sensor system becomes
    \begin{equation}
    \label{eq:state}
    |\psi(t)\rangle_s=\frac{1}{2}\left[\mathrm{e}^{-i \mathcal{E}_{\lambda}t}|\phi_+(\lambda)\rangle_s+\mathrm{e}^{i \mathcal{E}_{\lambda}t}|\phi_-(\lambda)\rangle_s\right], 
    \end{equation}
    with the eigenvectors $|\phi_\pm(\lambda)\rangle_s=\vert 0\rangle_s\pm\delta_\lambda\vert 1\rangle_s$. We calculate the exact expression for the normalized population in the state $\vert 0 \rangle_s$ of the pseudo-Hermitian sensor system, conditioned on the ancilla qubit state $\vert 0\rangle_a$ as follows
    \begin{equation}
        \label{eq:SB}
        S(\lambda,t)=\frac{1}{1+\mathcal{D}_\lambda\tan ^{2}(\Phi_{\lambda,t})},
    \end{equation}
    with $\mathcal{D}_\lambda=\delta_\lambda^2$ and $\Phi_{\lambda,t}=\mathcal{E}_{\lambda} t$. It can be seen that when $\Phi_{\lambda,t}\simeq\pi/2$, $S|_{\mathcal{D}_\lambda=0}=1$, while a small non-zero value of $\mathcal{D}_{\lambda}$ will result in a significant change of $ S|_{0<\mathcal{D}_\lambda\ll 1}\rightarrow 0$. Thus, the population signal in Eq.(\ref{eq:SB}) would be quite sensitive to the parameter $\lambda$ when $0<\mathcal{D}_{\lambda}\ll 1$ and $\Phi_{\lambda,t}\simeq\pi/2$, which can be viewed as a parametric analog of subwavelength "dark state" optical potentials \cite{Lacki2016,Wang2018}. Therefore, the system can serve as an efficient sensor for the determination of the parameter $\lambda$. 

   {\it Divergent feature of susceptibility.---} The most remarkable feature of the normalized state population $S(\lambda,t)$ is that it shows ultra-sharp dips if the eigenstates $|\phi_\pm\rangle_s$ nearly coalesce with each other (namely $0<\mathcal{D}_{\lambda}\ll1$). Such an observation is verified in Fig.\ref{fig:signal}, in which the population $S(\lambda,t)$ exhibits sharp dips with a narrow width $\Delta t=2\sqrt{\mathcal{D}_\lambda} /\mathcal{E}_\lambda$ when the evolution time $t=t_j$ nearly satisfying the condition $\Phi_j=\mathcal{E}_{\lambda} t_j=(j-1/2)\pi$ ($j\in\mathbb{Z}^{+}$). Both the location $t_j$ and the width $\Delta t$ of the dip depend on the parameter $\lambda$ as $(\partial t_j/\partial\lambda)/\Delta t\sim\varepsilon^{-1}\omega^{-1}$ and $(\partial \Delta t/\partial \lambda)/\Delta t\sim\varepsilon^{-3/2}\omega^{-1} $ for the measurement working point $\lambda/\omega+2\varepsilon\sim\varepsilon^{3/2} $ \cite{supplement}, which implies that a slight change of $\lambda$ can result in prominent dip shift and narrowing (or broadening) as compared to the dip width itself.  Therefore, one can expect that $S(\lambda,t)$ would become extremely sensitive to the parameter $\lambda$ if the evolution time is set close to $t_j(\lambda_0)$, where $\lambda_0$ represents an approximate value of the parameter $\lambda$.
As an illustrative example, we set the evolution time as $\tau=\pi/4\omega\sqrt{ \varepsilon(1+ \varepsilon)}$ and thus $S(\lambda,\tau)\vert_{\lambda=0}=0$. When $\lambda=-2\varepsilon\omega$, the two eigenvectors coalesce to $|0\rangle_s$, $\delta_{\lambda}=0$ making the spin always stay in the  state $\vert 0 \rangle_s $, namely $S(\lambda,\tau)\vert_{\lambda=-2\varepsilon\omega}=1$. To obtain the explicit form for the peak width $\Delta \lambda$, we consider the regime in which $|(\lambda/\varepsilon\omega)+2|\leq 2$, and get  the following result up to the second-order of $(\lambda/\varepsilon\omega)+2$ as
    \begin{equation}
         \label{eq:S}
         S(\lambda,\tau)\simeq\pi^{2} \alpha^{2} \varepsilon/\left[\pi^{2} \alpha^{2} \varepsilon+(\frac{\lambda}{\varepsilon\omega}+2)^{2}\right]
    \end{equation}
    with $\alpha=(1/8)(\lambda/\varepsilon\omega)^{2}-(\lambda/\varepsilon\omega)$. From this approximate formula, it is straightforward to find that the peak width is $\Delta \lambda \simeq 3 \pi \varepsilon^{3/2} \omega$. In addition, we can also derive the susceptibility $\chi_\lambda=\partial S/\partial \lambda$, which indicates how sensitive the population $S(\lambda,\tau)$ would response to the change of the parameter $\lambda$ under a given evolution time $\tau$. Detailed calculations show that when $\lambda=\lambda_o$ which satisfies $\lambda_o/\omega+2\varepsilon =2.72\varepsilon^{3/2} $, the maximum susceptibility $\left|\chi_\lambda\right|_{\text{max}}\simeq 0.14\varepsilon^{-3/2} \omega^{-1}\sim \varepsilon^{-1} \mathcal{E}^{-1}$ is obtained \cite{supplement}, exhibiting divergent behavior for an infinitesimal $\varepsilon$. Note that the divergent feature may appear far away from an exceptional point when the energy splitting ($\sim \mathcal{E}$) is a constant finite value. We remark that $\lambda=\lambda_o$ represents the optimal measurement working point for the estimation of the parameter $\lambda$. 
    \begin{figure}[t]
    \centering
    \includegraphics[width=86mm]{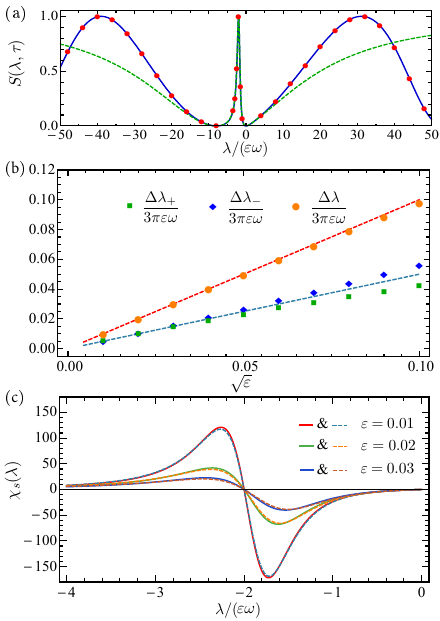}
    \caption{Pseudo-Hermitian quantum sensing with the dilated Hamiltonian in Eq.\eqref{eq:dilated_H} for the estimation of the parameter $\lambda$. (a) The normalized state population $S(\lambda,\tau)$ of a fixed evolution time $\tau=\pi/(2\mathcal{E})$ as a function of $\lambda/(\varepsilon \omega)$, shows a sharp peak located at $\lambda_m=-2\varepsilon\omega$. The derived analytical result in Eq.\eqref{eq:SB} (blue line) agrees well with numerical simulation (red circle). In the proximity of the peak,  the approximation of the state population in Eq.\eqref{eq:S} (green) allows us to derive the peak width $\Delta\lambda \simeq 3 \pi  \varepsilon^{3/2}\omega$. (b) The peak width as a function of the parameter $\varepsilon$, becomes ultra-sharp as $\varepsilon \rightarrow 0$. The full peak width $\Delta \lambda=\Delta\lambda_+ +\lambda_-$  (orange dots) is obtained by numerically solving $S(\lambda_m+\Delta\lambda_+)=S(\lambda_m-\Delta\lambda_-)=1/2$, which is well fitted by $\Delta \lambda = 3 \pi \varepsilon^{3/2}\omega$ (red dashed). (c) The susceptibility reaches the maximum value $\left|\chi_{s}(\lambda)\right|_{\text{max}} \simeq 0.14\varepsilon^{-3/2}\omega^{-1}$ which serves as the optimal measurement working point. The solid line is calculated exactly from Eq.\eqref{eq:SB} and the dashed line corresponds to an approximate analytical result, which enables us to derive information on the optimal measurement working point  see Ref. \cite{supplement}. The other parameters are $\omega = 1$, $\varepsilon=0.01$ and $\kappa = \delta$.}\label{fig:result}
    \end{figure}

   {\it Analysis of sensing performance.---} The experimental realization of a pseudo-Hermitian qubit sensor relies on the conditional evolution in a dilated Hermitian quantum system \cite{Tang2016,Xiao2018,Wu2019}. This approach results in  a finite success probability $\gamma(\lambda_o,\tau)\simeq  (7.4-10.4 \sqrt{\varepsilon}) \varepsilon^{2}$ \cite{supplement}, which equals to the probability of the ancilla qubit in the state $|0\rangle_a$, see Fig.\ref{fig:model}. As expected, there is a trade-off between the enhanced susceptibility and the decreased success probability when setting $\varepsilon\ll 1$.  We investigate this trade-off by invoking the quantum Fisher information (QFI) analysis. The normalized state of the pseudo-Hermitian sensor is
    \begin{equation}
       \frac{ \vert\psi(\tau)\rangle_s}{\left\lVert\psi(\tau)\right\rVert}={S^{1/2}(\lambda,\tau)}\vert 0\rangle-i\left[  {1-S(\lambda,\tau)} \right]^{1/2}\vert 1\rangle.
    \end{equation}In an experiment involving $N$ experiment runs, the QFI for all successful measurements is found to be $\mathcal{I}(\lambda)=\gamma N\chi_\lambda^2/[S\left(1-S\right)]$, which gives $\mathcal{I}(\lambda_o)\sim N\varepsilon^{-1}\omega^{-2}$ where $\lambda_o$ is the working point with the maximum susceptibility. As comparison, for the dilated two-qubit system, an upper bound of the QFI for the parameter $\lambda$ can be derived as $\mathcal{K}_\lambda=4N\tau^2\sim N\varepsilon^{-1}\omega^{-2}$ with $\tau\simeq\pi/4\omega\sqrt{\varepsilon}$ \cite{supplement}.  We stress that $\mathcal{I}(\lambda_0)\sim\mathcal{K}_\lambda$ albeit the small success probability of the postselection measurement, implying that this particularly small sub-ensemble of all measurement outcomes, which reproduces the pseudo-Hermitian dynamics, is actually sufficient to achieve the QFI of the same order as the total system. Furthermore,  our following analysis shows that, the pseudo-Hermitian qubit sensor can provide superior performance over conventional sensing method of Ramsey-type interferometry when considering realistic noise \cite{Jordan2014}.
 For each experiment run, we assume that the readout of the sensor system is accomplished by state-selective fluorescence \cite{Kim2011}, where $m_0$ and $m_1$ photons are collected corresponding to the state $|0\rangle_s$ and $|1\rangle_s$ respectively. Without loss of generality, we assume that $m_0>m_1$. The optical readout thus maps the sensor system state into random variables $n_j=S(\lambda)\mathcal{N}(m_0,\sigma^2)+(1-S(\lambda))\mathcal{N}(m_1,\sigma^2)+\xi_j$ denoting the photon number in the $j$-th experiment run, where $\mathcal{N}(m_s,\sigma^2)$ represents the Gaussian distribution with the mean value $m_s$ and the variance $\sigma^2$. Apart from Gaussian noise, we also take into the other realistic noises in experiment, e.g. time-correlated background photons and device fluctuations, which is denoted by $\xi_j$. Under these conditions, implementing an optimal estimator to achieve the Cram\'er-Rao bound would require detailed knowledge about noises, which is challenging for typical experiments \cite{Jordan2014}. More practically, we can choose an estimator as $\hat{S}=(1/N) \sum_{j}\left(n_{j}-m_{1}\right)/\left(m_{0}-m_{1}\right)$, treating each experiment run equally. With detailed analysis \cite{supplement}, we find that the mean value is $\langle\hat{S}\rangle=S(\lambda)$ and the variance is
    \begin{equation}
        \left(\delta S\right)^2=\frac{(1-S) S}{(\gamma N)}+\frac{\sigma^{2}}{(\gamma N m^{2})}+\frac{1}{(\gamma^2 N^{2}m^{2})}\sum_{i j}\left\langle\xi_{i} \xi_{j}\right\rangle, 
        \label{eq:variance}
    \end{equation}where $m=m_0-m_1$ defines the readout fluorescence contrast between the state $\vert 0\rangle_s$ and $\vert 1 \rangle_s $. In Eq.(\ref{eq:variance}), the first and second terms result from shot noise and  Gaussian noise respectively, both satisfying standard quantum limit. Whereas the last term leads to a constant value $\xi^2/m^2$ independent of $N$ for fully correlated background noises with $\left\langle\xi_{i} \xi_{j}\right\rangle= \xi^2$, which represents the inevitable noise that cannot be averaged out  by repetitive measurements \cite{Starling2009,Feizpour2011}. Therefore, under the condition of $\gamma N\gg1$, we obtain $\delta S\simeq \xi/m$. In Ramsey-type interferometry, the population signal can be written as $d(\lambda)=\left(1+\cos(2\lambda \tau)\right)/2$.  Similar analysis allows us to obtain the corresponding variance as $\delta d\simeq \xi/m$ when $ N\gg 1$. 
    Based on the above analysis of susceptibility and noise, we can estimate the measurement sensitivity in the limit of large $N$ repetitive experiment runs, which is quantified by the minimum parameter change that can be detected above the noise level, as follows
    \begin{equation}
        \delta \lambda_{\text{min}}^{(s)}\simeq\delta S/\left|\chi_\lambda\right|_{\text{max}}\sim (\xi/m) \left(\varepsilon\sqrt{\varepsilon}\right)\omega.
    \end{equation}
The Ramsey-type method yields $\delta \lambda_{\text{min}}^{(d)}=\delta d/(\partial d/\partial \lambda)\simeq \delta d\sqrt{\varepsilon}\omega\sim (\xi/m) \omega\sqrt{ \varepsilon}$ for the same measurement time $\tau \simeq \pi/4\omega\sqrt{\varepsilon}$. The ratio between $\delta \lambda_{\text{min}}^{(s)}$ and $\delta \lambda_{\text{min}}^{(d)}$ is thus given by ${\delta \lambda_{\text{min}}^{(s)}}/{\delta \lambda_{\text{min}}^{(d)}} \sim  \varepsilon$, which indicates that our proposal offers enhanced sensitivity than the conventional Ramsey-type interferometry when there exists noise that is not dependent on the number of repetitive measurements \cite{Hosten2008,Dixon2009}.
    \textit{Feasibility of experimental realizations.---} The proposal can be realized in a number of physical settings, which requires the engineering of the dilated two-qubit Hamiltonian in Eq.\eqref{eq:dilated_H}. As an example, it can be implemented with trapped ions such as $\tensor[^{171}]{\mathrm{Yb}}{^+}$ ions in a linear radio-frequency Paul trap, where tunable spin-spin couplings have been realized using multiple transverse motional modes \cite{Kim2011,Richerme2014}. Effective two-level spins are represented by $|F=1,m_F=0\rangle$ and $|F=0,m_F=0\rangle$ hyperfine ground states of  $\tensor[^{171}]{\mathrm{Yb}}{^+}$ ions with frequency splitting $\omega_0/2\pi\approx 12.6\mathrm{GHz}$, whose coherence time can exceed $1 s$ \cite{Kim2011}. Spin-dependent forces excite virtual phonons which mediate the required interacting Hamiltonian between two ions $H=B(I\otimes \sigma_x)+J( \sigma_y\otimes \sigma_y)$. The parameters $J/2\pi=10\mathrm{kHz}$ and $B/2\pi\leq 1 \mathrm{kHz}$ are achievable as reported in the experiments \cite{Richerme2014}, which lead to an interrogation time $\tau\simeq \pi/2J\approx 25\mu s$ (well within the coherence time) for a single experiment  run. We remark that the required Hamiltonian in Eq.\eqref{eq:dilated_H} can also be realized in other physical systems, such as solid-state spin system in diamond and superconducting qubit system. In particular, the implementation of non-Hermitian dynamics based on Naimark dilation has been achieved using nitrogen-vacancy center in diamond by selective microwave pulses \cite{Wu2019} and single-photon interferometric network \cite{Xiao2018}, which facilitates the experimental realization of the present proposal.

  \textit{Conclusion $\&$ Outlook.---} We present a strategy to realize pseudo-Hermitian sensing with a single-qubit probe by embedding it into a dilated two-qubit Hermitian system. The eigenstate coalescence leads to divergent susceptibility in the normalized state population of the pseudo-Hermitian probe system conditioned on the measurement outcome of the ancilla qubit. We illustrate the advantages of our sensing strategy under the influence of realistic noises in addition to the shot noise. The proof-of-principle experiment realization is feasible using trapped ions and solid-state spins among other physical systems. We emphasize that the eigenstate coalescence does not necessarily lead to degenerate eigenvalues, therefore our strategy does not rely on the existence of exceptional points \cite{supplement}. Therefore, the proposal is expected to provide a new way to exploit non-Hermitian physics for quantum sensing and quantum metrology using qubit systems. 

   {\it Acknowledgements.---} We thank Prof. Jan Wiersig for providing us very helpful comments. The work is supported by National Natural Science Foundation of China (11574103, 11874024, 11690030, 11690032), the National Key R$\&$D Program of China (2018YFA0306600), the Fundamental Research Funds for the Central Universities. H. L. is also supported by the Young Scientists Fund of the National Natural Science Foundation of China (Grant No. 11804110).\\

 \bibliography{reference}

\end{document}